\documentclass[aip,sd,amsmath,amssymb,reprint]{revtex4-1}

\usepackage{graphicx}
\usepackage{dcolumn}
\usepackage{bm}
\usepackage{color}

\begin{document}

\title{Fractal and Fractional SIS model for syphilis data}

\author{Enrique C. Gabrick} 
\email{ecgabrick@gmail.com}
\author{Elaheh Sayari} 
\author{Diogo Leonai Marques de Souza}
\affiliation{Graduate Program in Science, State University of Ponta Grossa,
84030-900, Ponta Grossa, PR, Brazil.}
\author{Fernando da Silva Borges}
\affiliation{Department of Physiology and Pharmacology, State University of New
York Downstate Health Sciences University, 11203, Brooklyn, NY, USA.}
\author{Jos\'e Trobia}
\affiliation{Department of Mathematics and Statistics, State University of Ponta Grossa, 84030-900, Ponta Grossa, PR, Brazil}
\author{Ervin K. Lenzi} 
\affiliation{Graduate Program in Science, State University of Ponta Grossa,
84030-900, Ponta Grossa, PR, Brazil.}
\affiliation{Department of Physics, State University of Ponta Grossa, 84030-900, Ponta Grossa, PR, Brazil}
\author{Antonio M. Batista}
\affiliation{Graduate Program in Science, State University of Ponta Grossa,
84030-900, Ponta Grossa, PR, Brazil.}
\affiliation{Department of Mathematics and Statistics, State University of Ponta Grossa, 84030-900, Ponta Grossa, PR, Brazil}

\begin{abstract}
This work studies the SIS model extended by fractional and fractal 
derivatives. We obtain explicit solutions for the standard and fractal 
formulations; for the fractional case, we study numerical solutions. 
As a real data example, we consider the Brazilian syphilis data from 2011 
to 2021. We fit the data by considering the three variations of the model. 
Our fit suggests a recovery period of 11.6 days and a reproduction 
ratio ($R_0$) equal to 6.5. By calculating the correlation coefficient 
($r$) between the real data and the theoretical points, our results suggest 
that the fractal model presents a higher $r$ compared to the standard 
or fractional case. The fractal formulation is improved when two different 
fractal orders with distinguishing weights are considered. This modification 
in the model provides a better description of the data and improves the 
correlation coefficient.
\end{abstract}

\maketitle

\begin{quotation}
Mathematical models are a powerful tool to understand, forecast, and 
simulate control strategies for disease spread. In mathematical 
epidemiology, one of the most successful is the compartmental. This model 
type stores individuals in compartments according to their infection status. 
In general, the flux among the compartments is described by ordinary 
differential equations that have high accuracy in reproducing real data. 
In the classical formulations, the host population is divided into 
compartments of Susceptible ($S$), Exposed ($E$), Infected ($I$), and 
Recovery ($R$). Combinations of these compartments lead to the SI, SIS, 
SIR, and SEIR models used to study the spread of different diseases. For 
sexual diseases, such as gonorrhoea or syphilis, the adequate model 
is the SIS. The SIS model describes diseases which not confer immunity 
after the recovery period.  In this work, we study extensions of the SIS 
model via non-integer differential operators, fractional and fractal. We 
consider syphilis data from 2011 to 2021, collected in Brazil. Our results 
show that the fractal order operator is more efficient than the fractional 
and the standard to fit the considered data. Therefore, our methodology 
can be extended for different models and diseases to obtain the best 
description of real data.
\end{quotation}


\section{Introduction}
After the pioneering work of Kermack and McKendrick~\cite{Kermack1927} in 
1927, many works have considered compartmental models~\cite{Keeling2008}. 
This type of model compartmentalises the population into groups according 
to the infection status, which can be Susceptible ($S$), Exposed ($E$), 
Infected ($I$), and Recovered ($R$)~\cite{Bjornstad2018}. The $S$ compartment 
is related to healthy individuals; $E$ with the infected individuals, but 
not yet infectious; $I$ with the infectious individuals; and $R$ with the 
individuals who acquire immunity, permanent or not~\cite{Batista2021}. 
The combination of these compartments leads to the SI~\cite{Allen1994}, 
SIS~\cite{Gray2011}, SIR~\cite{Cooper2020}, SIRS~\cite{Aguiar2008}, 
SEIR~\cite{Brugnago2020}, and SEIRS~\cite{Michele2022} models.

These models have been successfully applied in several contexts~\cite{Hethcote2000}, 
for example in diseases as gonorrhoea~\cite{yorke1976}, COVID-19~\cite{Manchein2020}, 
HIV~\cite{Dalal2008}, influenza~\cite{Dushoff2004}, dengue~\cite{Aguiar2009}, 
and others~\cite{Amaku2021, Scarpino2019, Olsen1990, Keeling1999, Altizer2006, Galvis2022, Machado2021}. 
In addition to the success in modelling real data, the compartmental models 
can be easily adapted to study generic situations~\cite{Bjornstad2020, Shea2020, Bjornstad2020b}. 
For instance, the SEIR model can be adapted to study the effects of two vaccination 
doses in a determined population~\cite{Gabrick2022}. The inclusion of multi-strain 
in a SIR model describes the data of dengue, and, due to seasonality and multi-strain, 
the solutions become complex, i.e., chaotic~\cite{Aguiar2011}. Including 
seasonality in an SEIRS model can lead to coexistence between chaotic and periodic 
attractors~\cite{Gabrick2023}. This coexistence is associated with tipping points, 
which depend on the control parameter. A tipping point was also found when the 
network topology is considered in SIS or SIR model~\cite{Ansari2021}. 
Despite its simplicity, the SIS model can generate rich solutions, such as Turing 
patterns, when spatial dynamics are included~\cite{Guo2019}.

Although we have many possibilities for compartmental models, the appropriate 
choice of model is made based on considerations consistent with the 
disease~\cite{AndersonMay}. Some diseases do not confer long-immunity in 
the infected individuals~\cite{Saka2014}, such as rota-viruses~\cite{Dian2021}, 
sexually transmitted~\cite{Lynn2004}, bacterial~\cite{Ghosh2006, Feng2020}, 
and other types of infections~\cite{Pang2019, Misra2011}. For these diseases, 
the appropriate model is the SIS model~\cite{Wu2023, Hethcote1984}. Considering 
a SIS model with variable population size, Hethcote and van den Driessche~\cite{Hethcote1995} 
obtained persistence, equilibrium, and stability thresholds. Their results 
suggest that combinations of disease persistence and death rate can cause 
a decrease to zero in the population size. Furthermore, the endemic point 
is asymptotically stable for some parameters. However, for other parameter 
ranges, Hopf bifurcation emerges. Gray et al.~\cite{Gray2012} studied the 
effects of environmental noise in a SIS model. They obtained explicit 
solutions of the stochastic version of the model and compared them with 
numerical solutions. First, they consider a two-state Markov chain, then 
generalise the results to a finite one. Additionally, they consider a 
realistic scenario by considering the parameters of {\it Streptococcus pneumoniae} 
spread in children. The stochastic version of this model was obtained in 
the previous work~\cite{Gray2011}. Gao and Ruan ~\cite{Gao2011}, reported 
a SIS patch model with variable coefficients. In this formulation, the 
authors investigated the human movement's influence on the spread of 
disease among patches. They performed numerical solutions to study the 
two patches' situation. Also, considering the two-patch SIS model, Feng et al. 
explored the stability and bifurcation~\cite{Feng2020}.

In an attempt to improve this model, extensions have been proposed, for 
instance, the stochastic version proposed by Gray et al.~\cite{Gray2011} 
or the inclusion of reaction-diffusion terms~\cite{Sun2020, Ge2017, Cui2016, Cai2015}. 
An extension that has been gaining much attention is the inclusion of 
fractional derivatives in the SIS model~\cite{Wang2023, Wu2023, Liu2019, Hassouna2018,Balzoti2020, Balzoti2021, Xia2009, Abuasad2019, Hoang2020, Liu2021}. 
Fractional calculus has been advanced in different fields as a powerful 
approach to incorporate different aspects with extensions of the differential 
operators to a non-integer order~\cite{Book01}. Fractional operators have 
been applied in several scenarios, such as anomalous diffusion~\cite{metzler2000random,Book01,evangelista2023introduction}, 
anomalous charge transport in semiconductors~\cite{uchauikin2013fractional}, 
chaos~\cite{zaslavsky2002chaos}, magnetic resonance imaging~\cite{lenzi2022fractional,magin2019capturing}, 
and electrical impedance~\cite{barbero2022time,bisquert2001theory}.  
In epidemiological models, fractional calculus has been used to extend 
the differential operators and, consequently, 
the models~\cite{Angstmann2021, Angstmann2016, Sene2020, Taghvaei2020, Almeida2018, Ahmad2020, Cai2022}, 
allowing us to obtain different behaviours connected to the different relaxation processes. 
An important aspect of fractional calculus is the memory effect~\cite{Li2018}. 
Due to the non-locality, memory, and extra degree of freedom, the fractional 
epidemic models are richer compared to the standard ones~\cite{Balzoti2020, Balzoti2021}.

Despite fractional models gaining much attention, less attention has been 
devoted to extending the epidemiological models with fractal derivatives. 
The fractal operators, which use the concept of fractal space~\cite{Arif2021}, 
have been applied in many situations, such as porous media~\cite{Brouers2018}, 
anomalous diffusion~\cite{Chen2006, Chen2010}, heat conduction~\cite{Wang2012}, 
dark energy~\cite{He2014}, Casimir effect~\cite{Cheng2013}, and others~\cite{He2018, Zhang2009}. 
Compared to standard calculus, which considers a continuous space-time, 
fractal calculus has been shown to be more accurate when fitting the experimental data~\cite{He2018}.

In this work, we study fractional and propose fractal 
extensions of the SIS model calibrated with syphilis data from 
Brazil (available on Ref.~\cite{dados}) from 2011 to 2021. 
We consider the SIS model due to its simplicity and adequate description 
of sexual disease transmission~\cite{Keeling2008}. However, other 
models can be employed to study the syphilis spread~\cite{Roy2016, Milner2010, Iboi2016}.
We consider the simplest form of the SIS model, i.e., 
without demographic characteristics. We made this simplification considering 
that birth and death rates are practically constant in the time range 
considered~\cite{IBGE}. This research is organised as follows.
We first present the standard SIS model (Section II), and after that, we 
analyse the extensions based on the fractional (Section III) and fractal 
(Section IV) differential operators. In the standard and fractal cases, 
we obtain analytical expressions. For the fractional case, we consider 
the numerical integrator Predictor-Evaluate-Corrector-Predictor (PECE)~\cite{diethelm2005}. 
Our results suggest that fractal calculus presents a higher correlation 
coefficient in fitting the real data. Finally, in Section V, we draw our 
conclusions.

\section{Standard SIS Model}

\begin{figure}[hbt]
\centering
\includegraphics[scale=0.45]{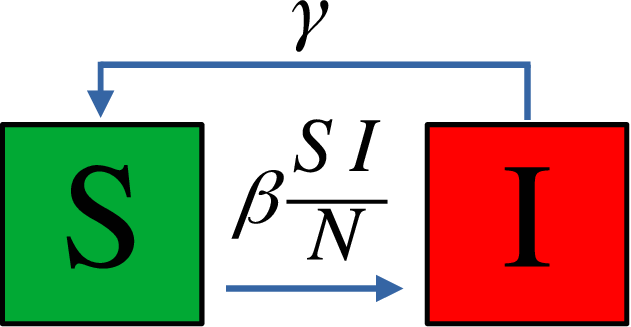}
\caption{Schematic representation of SIS model.}
\label{fig1}
\end{figure}

The SIS model compartmentalises the host population into Susceptible 
($S$) and Infected individuals ($I$)~\cite{Keeling2008}. The $S$ 
individuals are infected when in contact with $I$; after that, they 
evolve to the $I$ compartment by a transmission rate $\beta$. Once in 
the $I$ compartment, the individuals stay by an average time equal to 
$1/\gamma$, thenceforth they can be reinfected, as schematically 
represented in Fig.~\ref{fig1}.

The standard SIS model is described by the following equations~\cite{Batista2021}
\begin{eqnarray}
\frac{dS}{dt} &=& -\beta \frac{SI}{N} + \gamma I, \label{standard-1} \\
\frac{dI}{dt} &=& \beta \frac{SI}{N} - \gamma I, \label{standard-2}
\end{eqnarray}
{subject to the initial conditions $I(0) = I_0$ and $S(0) = S_0 = N - I_0$, 
where $N$, $\beta$, $\gamma$, $S_0$, and $I_0 \geq 0$~\cite{Moneim2005}. Therefore, 
for Eqs.~(\ref{standard-1}) and (\ref{standard-2}),
there exists only one solution for a given initial condition $S_0$, $I_0$ 
for all $t \geq 0$, defined in $D = \{(S, I) \in [0,N]^2 \ | \ S + I = N\}$. 
The proof of this statement is  straightforward using the techniques 
reported in~\cite{Hale1969}. In addition, as $S_0^+$ and $I_0^+$, we have 
$S(t)$ and $I(t) \geq 0$ for all $t \geq 0$. The proof is found in 
Refs.~\cite{Gao2008, Gao2011b}.

Equations~(\ref{standard-1}) and~(\ref{standard-2}) 
are population size dependent. To normalise it, it is necessary to impose the 
transformations $S \xrightarrow{} sN$ and $I \xrightarrow{} iN$, which are 
valid for constant population size. As we are considering data from 2011 
to 2021, we  normalise the model. 
According to Brazilian Institute of Geography and Statistics (IBGE, Portuguese abbreviation), 
the Brazilian population increased by around 0.52\% per year in the range 
2010 to 2022~\cite{IBGE, IBGE2}.
In this way, our assumption is reasonable 
and we can neglect demographic characteristics.}
With these transformations, the equations are given by
\begin{eqnarray}
\frac{ds}{dt} &=& -\beta si + \gamma i, \label{sis-eq1} \\
\frac{di}{dt} &=& \beta si - \gamma i, \label{sis-eq2}
\end{eqnarray}
subject to the initial conditions $i(0) = i_0$ and $s(0) = 1 - i_0$. 
The 
solutions with biological meaning are restricted to $s,i \in [0,1]$.
The sum of Eqs.~(\ref{sis-eq1}) and~(\ref{sis-eq2}) results in $s + i = 1$. 
With this constraint, Eq.~\ref{sis-eq2} is rewritten as
\begin{equation}
    \frac{di}{dt} = i \beta[1 - i - R_0^{-1}(t)],
    \label{eq1}
\end{equation}
where $R_0 = \beta/\gamma$ (Ref.~\cite{Keeling2008}). 

Let $i(0) = i_0$, by integration of Eq.~\ref{eq1}, we obtain
\begin{equation}
    i(t) = \frac{i_0 \xi e^{\xi \beta t}}{\xi + i_0 (e^{\xi \beta t} - 1)},
    \label{sol1}
\end{equation}
where $\xi \equiv (1- R_0^{-1})$ and $s(t)$ is immediately determined by 
$s(t) = 1.0 - i(t)$. In the limit $t \rightarrow \infty$, Eq.~\ref{sol1} 
results in $1 - R_0^{-1}$, which corresponds to the equilibrium state~\cite{Keeling2008}.

\begin{figure}[hbt]
\centering
\includegraphics[scale=0.8]{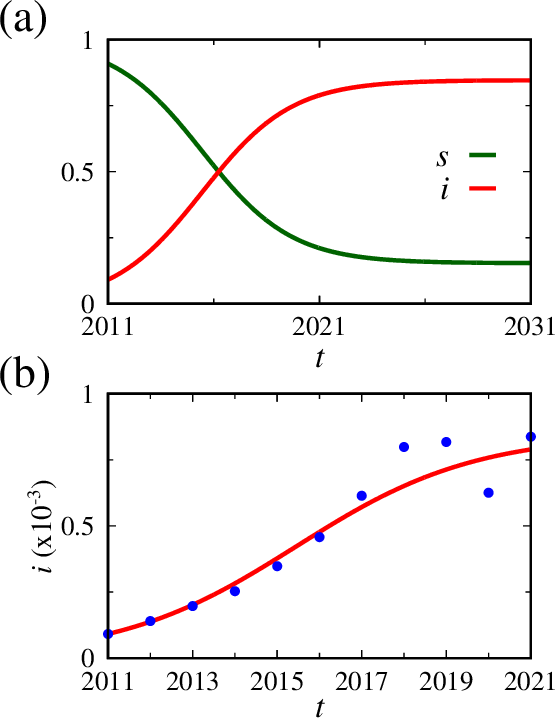}
\caption{(a) Solution of the SIS model. The green curve is related to 
$s$ and the red one to $i$. (b) Amplification of $i$ curve in red line 
and real data in blue points. The correlation coefficient is $r = 0.9900$.  
We consider $\beta = 204.4$, $\gamma = 31.39$, $R_0 = 6.5$, and $i_0 = 0.0912$. The 
population is normalised for $N=200$ million.}
\label{fig2}
\end{figure}

Considering the Brazilian data from syphilis available on Ref.~\cite{dados}, 
from 2011 to 2021, the best fit suggests $\beta = 204.4$, and 
$\gamma = 31.39$ (years$^{-1}$). 
To obtain the best fit, we first calibrate the model by the 
Levenberg–Marquardt non-linear least-squares algorithm in the R package 
minpack.lm~\cite{Elzhov}. After that, we compute the correlation coefficient ($r$) 
between real and simulated data in C language and consider the parameters 
which maximise $r$.
The time evolution of the model with these parameters is 
shown in Fig.~\ref{fig2}(a). These parameters correspond to $R_0 = 6.5$. 
The real data are not normalised. However, without loss of generality, it is 
possible to normalise the population in relation to $N = 200$ 
millions (value suggested by the fit). 
In this case, $i$ gives us information about 
the fraction of infected each year.
{As the initial condition, we choose the fraction of infected people in 
the population at the beginning of the spread, i.e., our initial condition is equal 
to the initial value $i_0 = 0.0912$.}
With the $\gamma$ estimated, the average recovery period is equal to 11.6 days, 
which agrees with syphilis characteristics. For these parameters, the 
correlation coefficient $r$ is equal to 0.9900. The $r$ value indicates a 
good fit, which can be seen in Fig.~\ref{fig2}(b) by the red line and the 
experimental points (blue points). The error associated with the explicit 
solution and the numerical integration (using the Runge-Kutta 4th order method) 
is $10^{-6}$. This model is very good at fitting the points until 2016. 
From this point, the theoretical model diverges from the experimental points.
\section{Fractional SIS model}
{To improve the fit of real data, we fix all the parameters 
found and varied the order of the differential operators to observe if 
improvement is obtained. As we are dealing 
with fractions of infected, i.e., normalised population, we can apply 
the extension directly in the system described by Eqs.~\ref{sis-eq1} and 
\ref{sis-eq2}}.
This extension is made by the replacement 
$\frac{d}{dt} \rightarrow{} \frac{d^\alpha}{dt^\alpha}$~\cite{evangelista2023introduction}.
In this work, we consider the Caputo fractional operator~\cite{Book01}, defined by
\begin{equation}
\frac{\partial^{\alpha}}{\partial t^{\alpha}}f (\vec{r},t)=
\frac{1}{\Gamma\left(1-\alpha\right)}
\int_{0}^{t}dt'\frac{1}{(t-t')^{\alpha}}\frac{\partial}{\partial t}f (\vec{r},t),
\label{caputo}
\end{equation}
where $\Gamma(\cdot{})$ is the Gamma function, and $\alpha \in (0,1)$. 
If $\alpha=1$, we recover the usual operator (standard case). Considering 
the Caputo operator, Eqs.~(\ref{sis-eq1}) and~(\ref{sis-eq2}) become
\begin{eqnarray}
\frac{d^{\alpha}s}{dt^{\alpha}} &=& -\beta si + \gamma i, \label{frac1} \\
\frac{d^{\alpha}i}{dt^{\alpha}} &=& \beta si - \gamma i, \label{frac2}
\end{eqnarray}
{defined in  $D = \{(s, i) \in [0,1]^2 \ | \ s + i = 1\}$.  
Given the initial condition $s_0$ and $i_0 \geq 0$, Eqs. (\ref{frac1}) 
and (\ref{frac2}) admit a unique solution $s(t)$ and $i(t)$~\cite{Hassouna2018}, which are positive for all $t \geq 0$~\cite{Balzoti2021}.}

Considering $s+i=1$, Eqs.~(\ref{eq1}) and~(\ref{sis-eq2}) are rewritten as
\begin{equation}
    \frac{d^{\alpha}i}{dt^{\alpha}} = i \beta[1 - i - R_0^{-1}].
    \label{eq1-frac}
\end{equation}

\begin{figure}[hbt]
\centering
\includegraphics[scale=0.85]{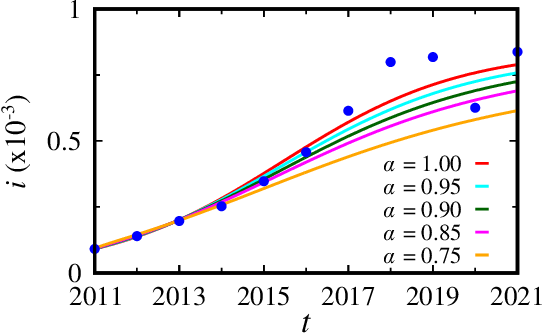}
\caption{Solutions for the fractional SIS model. The red line is for 
$\alpha=1$ ($r = 0.9900$), cyan line for $\alpha=0.95$ ($r = 0.9891$), 
green line for $\alpha=0.90$ ($r = 0.9880$), magenta line for $\alpha=0.85$ 
($r = 0.9868$), and orange line for $\alpha=0.75$ ($r = 0.9834$). 
We consider $\beta = 204.4$, $\gamma = 31.39$, $R_0 = 6.5$, and $i_0 = 0.0912$.}
\label{fig3}
\end{figure}

Due to the nature of fractional operators and the nonlinear aspect of 
the previous equation, an analytical expression for Eq.~(\ref{eq1-frac}) 
as in the previous section is not possible. Numerical solutions are 
feasible and can be found using the PECE method~\cite{diethelm2005}. 
{ As we are working with Caputo's definition, the $R_0$ is equal 
to $\beta/\gamma$ and, as all parameters are positive, the solutions 
stay positive respecting $1 = s + i$.} 
Numerical solutions are shown in Fig.~\ref{fig3} for $\alpha=1$ (red line), 
$\alpha=0.95$ (cyan line), $\alpha=0.90$ (green line), $\alpha=0.85$ 
(magenta line), and $\alpha=0.75$ (orange line). The respective correlation 
coefficients are $r = 0.9900$, $r = 0.9891$, $r = 0.9880$, $r = 0.9868$, 
and $r = 0.9834$. Note that $r$ decreases as a function of $\alpha$. In 
this way, no improvement in the fit occurs when fractional operators are 
considered. This happens because of the nature of the data points. The 
points increase after 2016, and the fractional derivative slows down the 
$i$ curve~\cite{Balzoti2020}. This effect can simulate a control measure 
in which memory effects are embedded.
\section{Fractal model}
As the next extension of the model, we consider the fractal derivatives, 
given by the following definition 
\begin{equation}
    \frac{df}{dt^\alpha} = \frac{1}{\alpha}t^{1-\alpha}\frac{df}{dt}.
    \label{fractal}
\end{equation}
This definition is known as Hausdorff derivative~\cite{He2018}.

For extending the Eqs.~(\ref{sis-eq1}) and (\ref{sis-eq2}) 
to fractal order, we apply a direct substitution of Eq.~(\ref{fractal}) 
into the Eqs., and obtain
\begin{eqnarray}
    \frac{ds}{dt} &=& \alpha t^{\alpha-1}\big(-\beta si + \gamma i \big), \label{fractal-eq1} \\
    \frac{di}{dt} &=& \alpha t^{\alpha-1}\big(\beta si - \gamma i \big), \label{fractal-eq2}
\end{eqnarray}
{where $\alpha>0$. Due to the direct connection of fractal derivative 
with standard one and $t \geq 0$, all the assumptions made for 
the positive solutions of Eqs.~(\ref{standard-1}) and (\ref{standard-2}) 
remain valid for Eqs.~(\ref{fractal-eq1}) and (\ref{fractal-eq2}).}
As all parameters are positive, including $t$, the solutions 
stay preserve $1 = s + i$. 
Similarly to Eq.(~\ref{eq1}),
\begin{equation}
    \frac{di}{dt} = \alpha t^{\alpha - 1} i \beta[1 - i - R_0^{-1}].
    \label{fractal-eq}
\end{equation}
An explicit solution for Eq.~(\ref{fractal-eq}) is possible and is given by
\begin{equation}
    i(t) = \frac{i_0 \xi e^{\xi \beta t^\alpha}}{\xi + i_0 (e^{\xi \beta t^\alpha} - 1)},
    \label{fractal-sol1}
\end{equation}
where $\xi \equiv (1- R_0^{-1})$, and $\alpha>0$. {The new 
parameter $\alpha$ is a multiplicative constant, then $R_0 = \beta/\gamma$.}

\begin{figure}[hbt]
\centering
\includegraphics[scale=0.8]{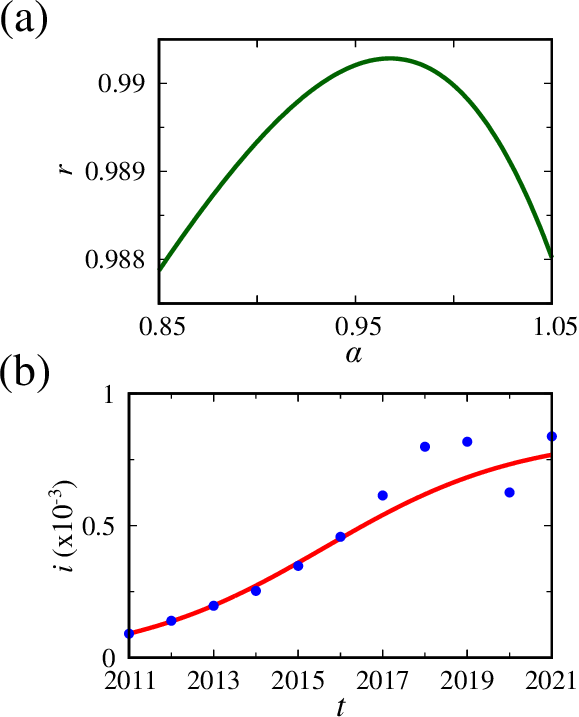}
\caption{Correlation coefficient as a function of $\alpha$ in the panel (a). 
Time series for $i$ in the panel (b) for $\alpha = 0.9673$, in red line 
and experimental data in blue points, with $r = 0.99028$.   
We consider $\beta = 204.4$, $\gamma = 31.39$, $R_0 = 6.5$, and $i_0 = 0.0912$.}
\label{fig4}
\end{figure}

As a new degree of freedom is included in the model, we expect a better 
fit of the data set. Figure~\ref{fig4}(a) displays $r$ as function of 
$\alpha$. Differently from the fractional case, the fractal derivative 
exhibits one point which maximises $r$. This point is $\alpha=0.9673$ and 
$r=0.99028$. Considering this $\alpha$ value, the solution for the model 
is shown in Fig.~\ref{fig4}(b) by the red line. With this extension, the 
theoretical model reproduces the data with more precision when compared to the standard 
and fractional cases. The standard case is recovered when $\alpha \rightarrow1.$

Although we improved the fit by including the fractal derivative, the points 
from 2017 remain away from the red curve. Considering the improvement 
given by the fractal derivative, we hypothesise that the experimental 
data are dominated by one fraction, namely $\sigma_1$, with weight $a_1$ 
in a certain range of time and, after that, by other fraction $\sigma_2$, 
with the weight $a_2$. To include these modifications, we consider 
$t^\alpha \rightarrow a_1 t^{\sigma_1} + a_2 t^{\sigma_2}$, obtaining 
the expression
\begin{equation}
    i(t) = \frac{i_0 \xi e^{\xi \beta (a_1 t^{\sigma_1} + a_2 t^{\sigma_2})}}{\xi + i_0 (e^{\xi \beta (a_1 t^{\sigma_1} + a_2 t^{\sigma_2})} - 1)},
    \label{fractal-sol2}
\end{equation}
where $\sigma_{1,2} > 0$ and $a_{1,2}$ are real positive constants.

\begin{figure}[hbt]
\centering
\includegraphics[scale=0.85]{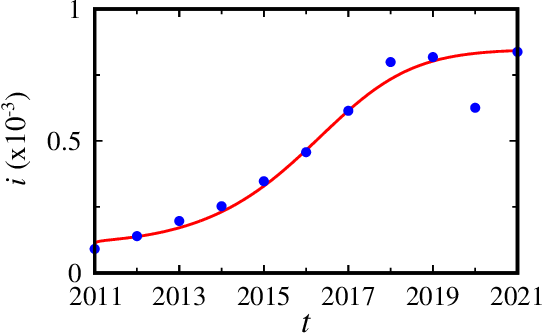}
\caption{Time series of $i$ in the red line and real data in blue points. 
We consider $\beta = 204.4$, $\gamma = 31.39$, $a_1 = 0.184$, $\sigma_1 = 1.9$, 
$a_2 = 0.82$, and $\sigma_2 = 0.1$. The parameter adjustment is $r=0.9980$.}
\label{fig5}
\end{figure}

Fixing $\beta = 204.4$ and $\gamma = 31.39$, the best fit is given for 
$a_1 = 0.184$, $\sigma_1 = 1.9$, $a_2 = 0.82$, and $\sigma_2 = 0.1$, as 
can be seen in Fig.~\ref{fig5}. Our hypotheses result in $r=0.9980$. The 
inclusion of two fractal orders modulated the curve behaviour in the different 
time ranges. For example, for the selected parameters, if we change $a_1$ 
the curve shape changed after 2016. In this way, the first part ($t>2016$) 
is dominated by $\sigma_1$. On the other hand, if we increase or decrease 
$a_2$, the first half of the curve changes for the selected parameters. 
Due to this characteristic, the SIS model is improved to fit the syphilis 
Brazilian data. The point located in 2020 diverges from the behaviour, 
which is not considered in the fit data.

\begin{figure}[hbt]
\centering
\includegraphics[scale=0.85]{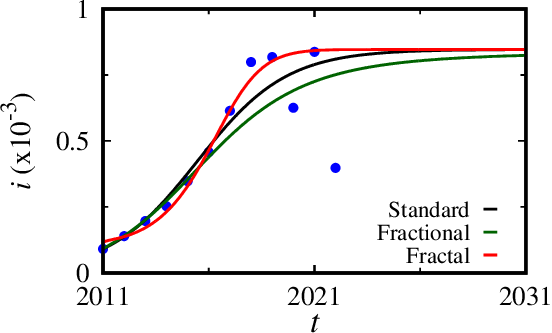}
\caption{Comparison among the three models versus data points.
Standard case (black line), the fractional case with $\alpha=0.9$ 
(green line), and fractal situation with $a_1 = 0.184$, $\sigma_1 = 1.9$, 
$a_2 = 0.82$, and $\sigma_2 = 0.1$ (red curve). We consider
$\beta = 204.4$, $\gamma = 31.39$, $R_0 = 6.5$, and $i_0 = 0.0912$.}
\label{fig6}
\end{figure}

Figure~\ref{fig6} shows the extension of the $i$ values overtime  
for the standard case (black line), fractional case with $\alpha=0.9$ 
(green line), and fractal situation with $a_1 = 0.184$, $\sigma_1 = 1.9$, 
$a_2 = 0.82$, and $\sigma_2 = 0.1$ (red curve). 
Superposing these solutions, 
we show that the fractal model reproduces the data with more precision. 
{Table 1 shows, for the range 2011-2021, the Mean Absolute Error 
(MAE) equal to 0.05, 0.06, and 0.03 for the standard, fractional, and fractal 
approaches, respectively. Considering the last 
point, 2022, the respective MAE changes to 0.09, 0.08, and 0.06. Therefore, 
our results suggest that the best model to describe the considered data 
is the fractal one. The last point increases the error due to the fact that 
we do not employ control measures. This point is in 2022 and can be associated 
with social behaviour changes or measures of error during the pandemic~\cite{Wu2023}. 
To better fit the social behaviour, it is necessary to change $\beta$ or the 
respective fractional order. As our goal is to reproduce and explains the 
previous data, we do not take into account control measures, which will be 
considered in future works. }

\begin{table*}[htb]\label{tab1}
\begin{tabular}{|c|c|c|c|c|c|c|c|c|c|c|c|c|c|c}
\hline
$\times10^{-3}$& 2011   & 2012    & 2013   & 2014   & 2015   & 2016   & 2017   & 2018   & 2019   & 2020   & 2021   & 2022   & {\bf MAE} 11-21 (22)  \\
\hline
Data           & 0.0912 & 0.1397  & 0.1966 & 0.2530 & 0.3476 & 0.4575 & 0.6142 & 0.7986 & 0.8176 & 0.6257 & 0.8376 & 0.3979 &           		  						\\
\hline
Standard       & 0.0912	& 0.1375  &	0.2011 & 0.2824 & 0.4771 & 0.5712 &	0.6511 & 0.7132 & 0.7582 & 0.7893 & 0.8099 & 0.8233 &           		  					\\
\hline
Error          & 0	    & 0.0022  &	0.0045 & 0.0294 & 0.1295 & 0.1137 &	0.0369 & 0.0854 & 0.0594 & 0.1636 &	0.0277 & 0.4254 & {\bf 0.05} ({\bf 0.09})			\\
\hline
Fractional     & 0.0912 & 0.1407  & 0.2009 & 0.2735 & 0.3549 & 0.4387 & 0.5181 & 0.5878 & 0.6452 & 0.6905 & 0.7251 & 0.7511 &           		  						\\
\hline
Error          & 0	    & 0.0010  &	0.0043 & 0.0205 & 0.0073 & 0.0188 & 0.0961 & 0.2108 & 0.1724 & 0.0648 & 0.1125 & 0.3532 & {\bf 0.06} ({\bf 0.08})			\\
\hline
Fractal        & 0.0912 & 0.1377  &	0.1712 & 0.2315 & 0.3290 & 0.4651 &	0.6147 & 0.7340 & 0.8022 & 0.8318 &	0.8422 & 0.0845 &           		  						\\
\hline
Error          & 0	    & 0.0020  &	0.0254 & 0.0215 & 0.0186 & 0.0076 &	0.0005 & 0.0646	& 0.0154 & 0.2061 &	0.0046 & 0.4474 & {\bf 0.03} ({\bf 0.06})      \\
\hline    
\end{tabular}
\caption{Fraction of syphilis case per year followed by the predicted value 
with respective absolute error. In the last column, there is the Mean 
Absolute Error (MAE) computed from 2011 until 2021 and 2011 until 2022.}
\end{table*}

\begin{figure*}[hbt]
\centering
\includegraphics[scale=0.6]{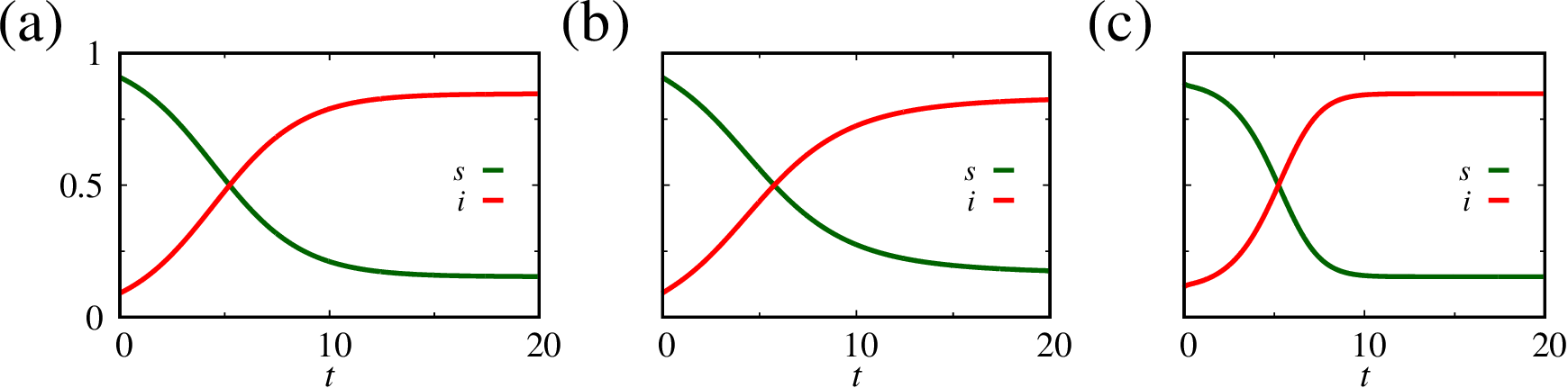}
\caption{Numerical solutions for $s$ (green line) and $i$ (red line). 
Panel (a) is for the standard case, the panel (b) is for the fractional case, with 
$\alpha = 0.9$, and the panel (c) for fractal case with $\sigma_1 = 1.9$, 
$a_2 = 0.82$, and $\sigma_2 = 0.1$. We consider
$\beta = 204.4$, $\gamma = 31.39$, $R_0 = 6.5$, and $i_0 = 0.0912$.}
\label{fig7}
\end{figure*}

{Finally, Fig.~\ref{fig7} displays generic solutions for 
the standard model in panel (a), for the fractional model in  
panel (b) (with $\alpha=0.9$), and for the fractal model in panel 
(c) ($\sigma_1 = 1.9$, $a_2 = 0.82$, and $\sigma_2 = 0.1$). We consider 
$\beta = 204.4$ and $\gamma = 31.39$. The green curve is related to $s$ 
and red one with $i$ solutions. From this solution, we note that the 
fractional situation takes more time to reach a steady solution than 
fractal and standard formulations. On the other hand, the fractal formulation 
reaches the steady solution faster than the other cases, as shown in panel (c).}
\section{Conclusions}
In this work, we considered the SIS model without demographic characteristics 
and analysed the extensions described by the substitution of integer operators 
by non-integer (fractional and fractal). We obtained analytical solutions 
for the standard (i.e., integer order) and the fractal case. Regarding 
the fractional situation, we studied the numerical solutions.  

Considering these three formulations, we investigated real data from 
Brazilian syphilis. From 2011 to 2021, our simulations 
show a basic reproduction number equal to 6.5. We
calculated the correlation coefficient between the experimental and
theoretical points, namely $r$, to measure the best fit. For the standard case, we obtained $r=0.99$. 
This formulation adjusts the real data with a good approximation. However, 
after a specific time (2016), the points follow a different trajectory 
than the predicted by the model. To adjust these points, we first hypothesise the fit by 
fractional derivatives due to the increase in the degree of freedom. Our 
results showed a slowdown in the infected curve, which followed an opposite 
behaviour compared with points after $t=2017$. The $r$ increases when the 
fractional order ($\alpha$) tends to the unity, i.e., recovering the 
standard case. In this situation, it was not possible to improve the fit. 
The third consideration was the replacement of integer operators with 
fractal operators. In this case, we constructed a curve of $r$ as a function 
of $\alpha$. The curve has a maximum point in $\alpha=0.9673$. Therefore, 
the $r$ value is improved when fractal derivatives are considered. In this 
case, we obtain $r=0.99028$. Looking at the data behaviour, we observed a 
different increase in the years 2016 and 2017. In light of this characteristic, 
we hypothesised that the curve is dominated by one fractal order in a specific 
range of time and by another fractal order after this time. In this way, we 
considered two different fractal orders, and our hypothesis was confirmed. 
We obtain a correlation coefficient equal to 0.998. 
The fractal model with two orders described the data set 
with more accuracy than the other considered approaches. This result remains 
valid when uncertainty, a type of random noise, is added in the data set. 

The fractional and fractal operators are a simple way 
of extending the standard approach and incorporating different effects 
such as memory effects, long-range correlations, etc. These effects may 
be related to the relaxation processes present in the system, which deviates 
from Debye's case, characterised by exponential relaxations. The non-Debye's 
cases present a different behaviour, such as power-law, stretched exponential,  
mixing between these behaviours, among others. Thus, extending standard 
operators to fractional or fractal operators is a possibility of capturing 
these behaviours, which are unsuitable for the usual approaches. 
Our results show that the fractal formulation describes the 
data with great accuracy. Comparing our results to the other 
work~\cite{Silva2022}, employing geo-processing techniques, we 
conclude that in addition to the simplicity of the fractal model, it describes the 
data with a very well accuracy. Models considering more sophisticated statistical 
analysis and machine learning techniques are found in 
Refs.~\cite{Teixeira2023, Cervantes2020, Zhu2022}. 
In addition, our model describes the data with great accuracy 
in the range from 2011 to 2021. In 2022, there is a decrease in the syphilis 
cases, which is predicted only by the fractional formulation. Data of 2023 
are not available from official agencies, however, some brazilian regions 
report an increase, which is in agreement with our model.


\section*{Acknowledgements}
The authors thank the financial support from the Brazilian Federal Agencies (CNPq); 
CAPES; Funda\-\c c\~ao A\-rauc\'aria. S\~ao Paulo Research Foundation 
(FAPESP 2022/13761-9). E.K.L. acknowledges the support of the CNPq 
(Grant No. 301715/2022-0). E.C.G. received partial financial support from
Coordenação de Aperfeiçoamento de Pessoal de Nível Superior - Brasil (CAPES) 
- Finance Code 88881.846051/2023-01. We would like to thank  www.105groupscience.com.

\section*{DATA AVAILABILITY}
The data that supports the findings of this study are available within the article.


\end{document}